\documentclass[twocolumn,showpacs,preprintnumbers,amsmath,amssymb]{revtex4}

\usepackage{graphicx}
\usepackage{dcolumn}
\usepackage{bm}

\begin{document}

\preprint{}

\title{Sink flow deforms the interface between a viscous liquid and air \\ into
    a tip singularity
               }

\author{S. Courrech du Pont}
\author{J. Eggers}%

\affiliation{%
School of Mathematics, 
University of Bristol, University Walk, \\
Bristol BS8 1TW, United Kingdom 
}%

\date{\today}

\begin{abstract}
In our experiment, an interface between a viscous liquid and air is
deformed by a sink flow of constant flow rate to form a sharp tip.
Using a microscope, the interface shape is recorded down to a tip
size of 1 $\rm{\mu m}$. The curvature at the tip is controlled by the
distance $h$ between the tip and the sink. As a critical distance
$h^{\star}$ is approached, the curvature diverges like
$1/(h-h^{\star})^3$ and the tip becomes cone-shaped. As the distance
to the sink is decreased further, the opening angle of the cone
vanishes like $h^2$. No evidence for air entrainment was found,
except when the tip was inside the orifice.
\end{abstract}

\pacs{Valid PACS appear here}
\maketitle
Hydrodynamics very often leads to the spontaneous formation of 
very small structures, such as in the separation of a liquid drop from 
a nozzle \cite{E93,RRR01}, drop coalescence \cite{ELS99}, 
shock waves \cite{LL84}, free-surface cusps \cite{E01,LRQ03},
or tips \cite{T34,CN02,Ch04}. Mathematically, all the above phenomena 
correspond to a singularity \cite{E98,K97} of an underlying hydrodynamic
equation, in which the hydrodynamic fields become non-smooth.
Hydrodynamic singularities can be classified into two groups.
In the first group of 'true' singularities the production of small 
scales is cut off only by a microscopic length, such as the size of 
a molecule. In the second group, there exists some hydrodynamic 
mechanism which is able to 'regularize' the singularity.
In the past, all singularities which persist for a finite amount of 
time were found to belong to the second group, typical 
regularizing mechanisms being diffusion \cite{LL84}, 
surface tension \cite{JM92}, or the presence of an outer liquid \cite{LRQ03}. 
By contrast, in this letter we describe a stationary 
tip whose typical size is not limited by any hydrodynamic mechanism,
such as surface tension. 

Tips were first investigated by G.I. Taylor \cite{T34},
by placing a liquid drop of small viscosity in a viscous flow 
produced by counter-rotating rollers, stretching the drop. 
At its ends, the drop develops tips which become quite sharp, 
yet it has never been established whether the tip remains rounded 
on some scale \cite{S94}. Tips have recently received intense 
scrutiny \cite{G98,CN02,C04} owing to their potential in producing 
extremely thin jets, leading to numerous microfluidic 
applications \cite{G98,GG01,SSA04}.
In the viscous withdrawal experiment \cite{CN02,C04} a liquid is 
withdrawn through a straw placed above a liquid-liquid interface,
producing a sharp ``hump''. While the curvature of the hump appears to
diverge as the flow rate is increased, the (apparent) 
singularity is actually cut off at a typical size of 
$0.2 mm$ \cite{C04}, and the hump transforms into a jet.

In the present experiment we replace one of the liquids by a gas, 
finding no sign of the disappearance of the tip. Hence we
are able to track the approach to the singularity (i.e. a 
divergence of the tip curvature) over four orders 
of magnitude, the curvature being described by a universal scaling law.
Three-dimensional axisymmetric 
flow thus appears to be fundamentally different from a two-dimensional 
one \cite{JM92}, in that surface tension does not succeed in regularizing the 
tip singularity. The presence of a singularity also raises exciting 
prospects to produce structures by hydrodynamic forcing whose size is only 
limited by microscopic features \cite{Z04}.
\begin{figure}
\includegraphics[width=5cm]{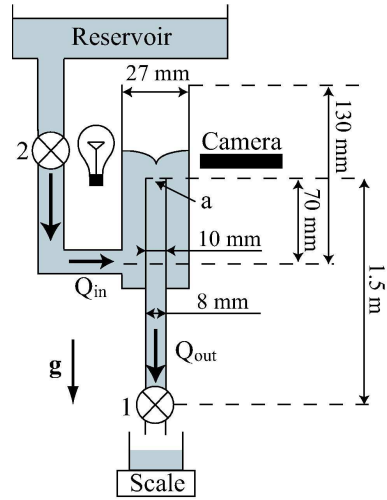}
\caption{Sketch of the experimental setup.\label{fig:shema}}
\end{figure}
\begin{figure}
\includegraphics[width=4.25cm]{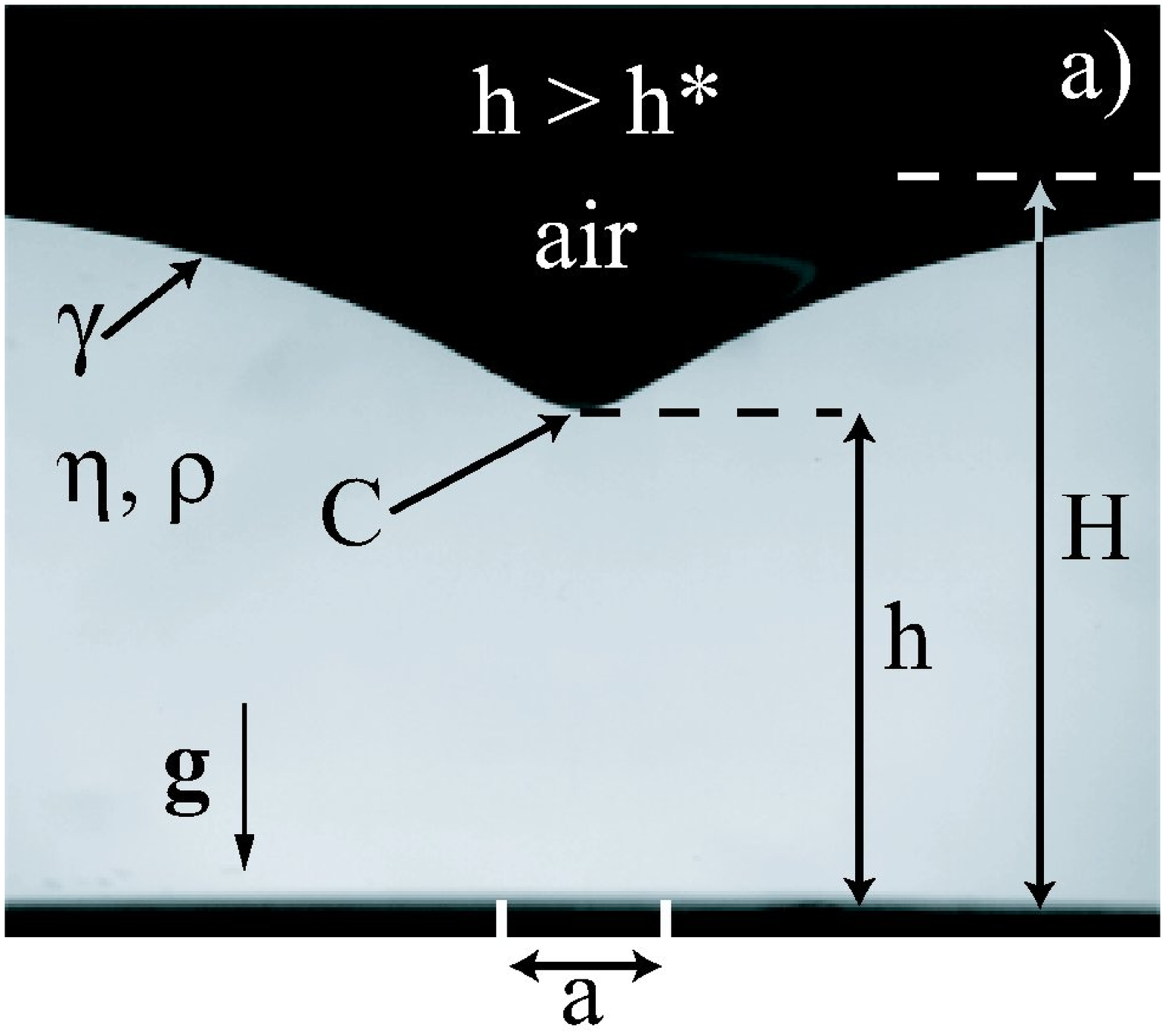}
\includegraphics[width=4.25cm]{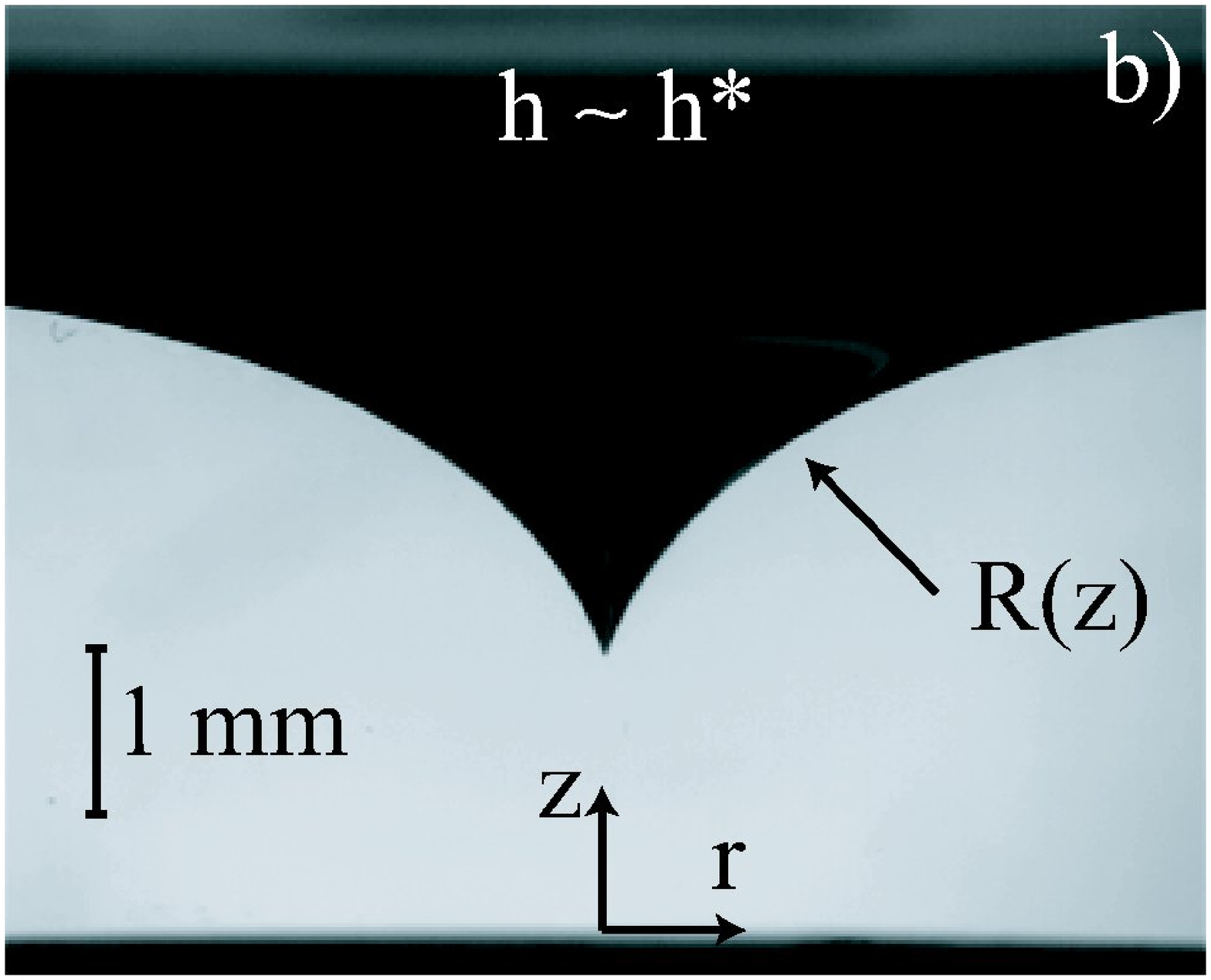}
\includegraphics[width=4.22cm]{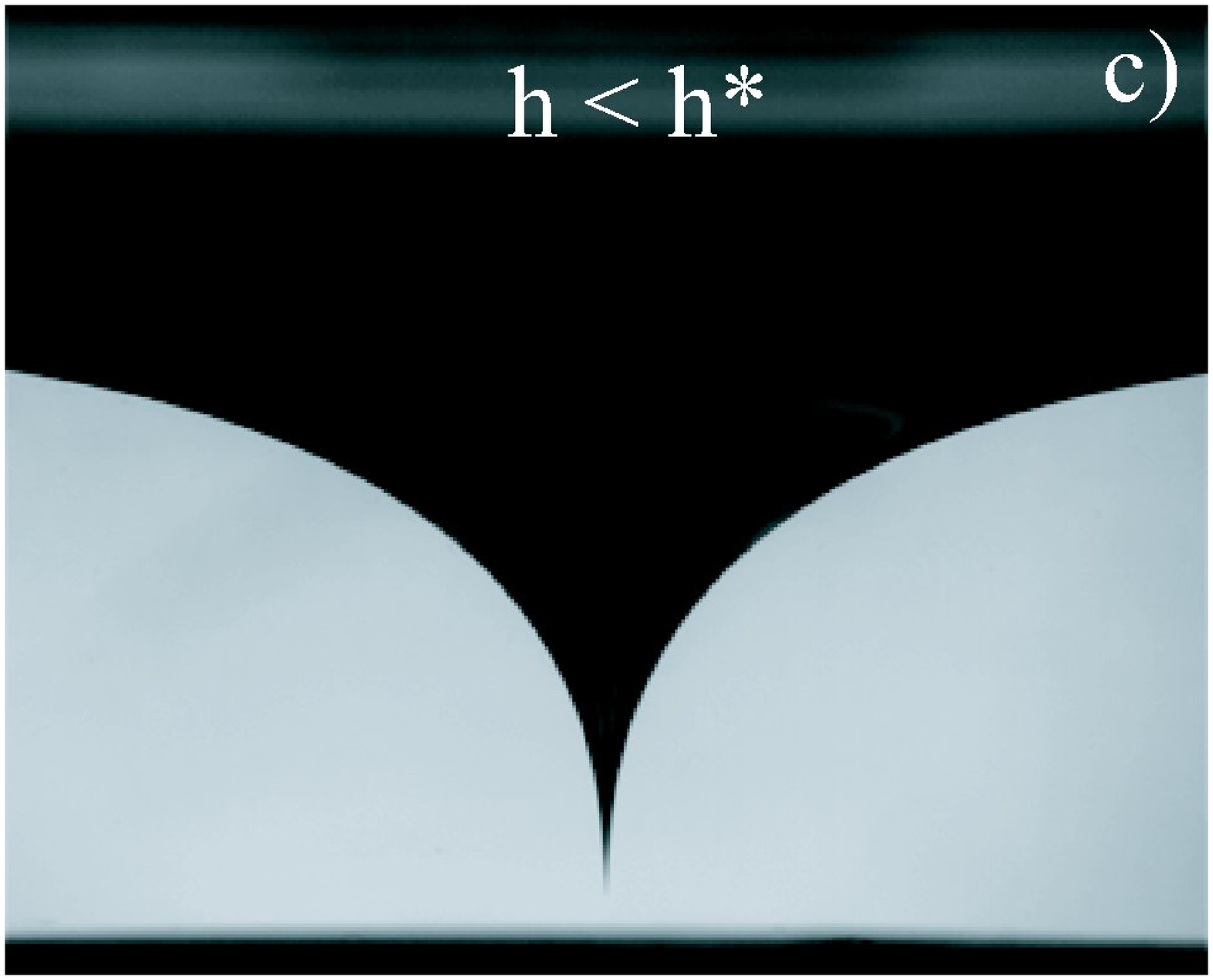}
\includegraphics[width=4.28cm]{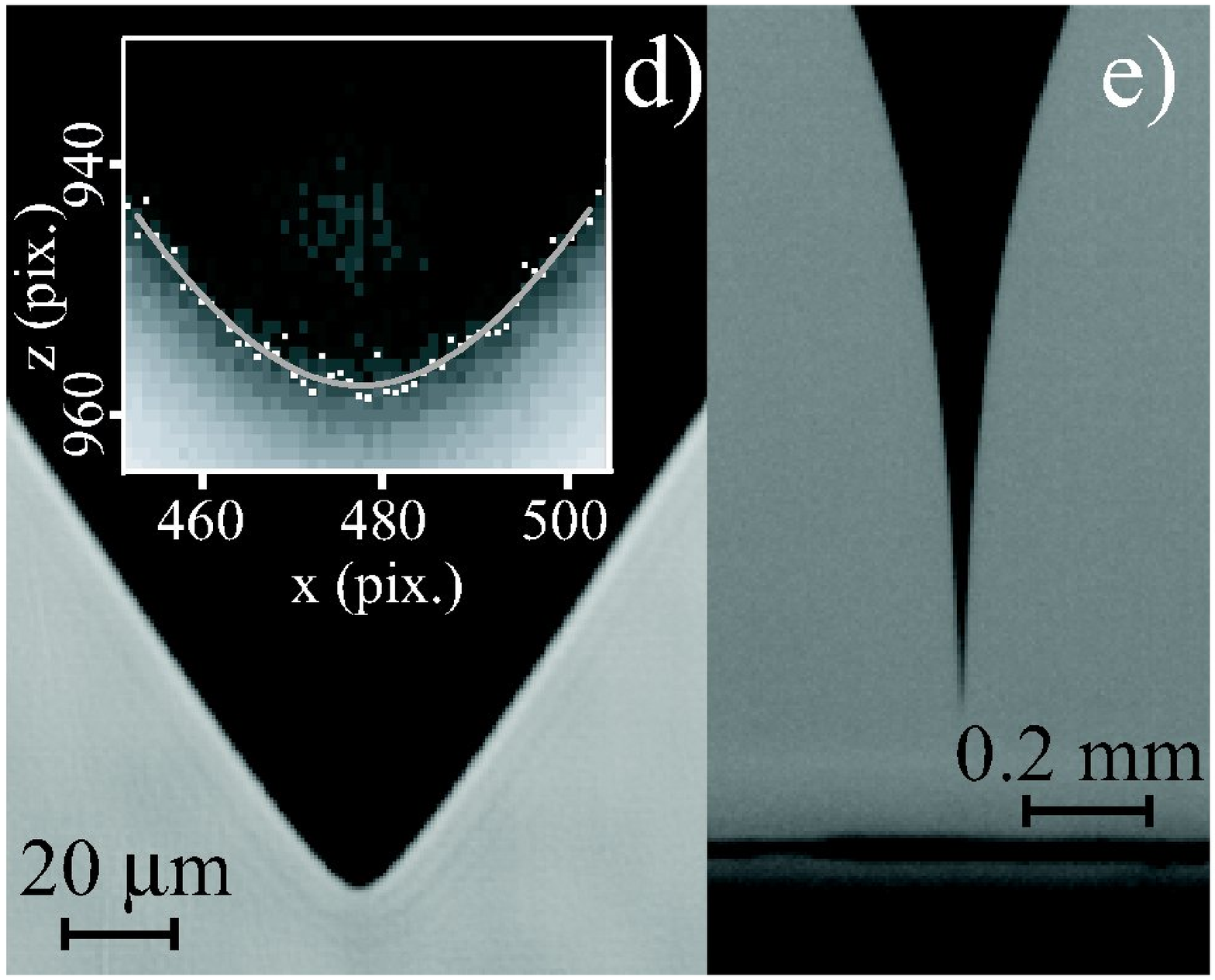}
\caption{Photographs of the air-liquid interface (silicone oil, 
$\eta$ = 60 Pa.s and $Q$= 3.9 $10^{-3}$ ml.s$^{-1}$). a): $h>h^{\star}$,
b): $h=h^{\star}$, c): $h<h^{\star}$.
d): a Gaussian fit to the tip just before the singularity.
e): the tip as it is about the enter the orifice.
\label{fig:image}}
\end{figure}

Our experimental setup, sketched in Fig.~\ref{fig:shema}, consists
of a $27\times27\times130$ mm cell partially filled with a
viscous liquid (we used castor oil and two different silicone oils). 
The liquid flows out of a hole ($a=1$ and 3 mm
in diameter) drilled into a circular plate of 0.7 mm thickness.
The flow rate $Q\equiv Q_{out}$ is adjusted to a constant value by 
the ball-valve (1), and measured by weighing. To ensure stationary 
conditions, the flow rate $Q_{in}$ into the cell is adjusted 
by the valve (2) to be almost equal but smaller than $Q_{out}$. 
In the course of a typical 
experiment, the cell empties over a period of 2-6 hours at constant $Q$,
thus adiabatically changing the filling height $H$ of the container. 

All results reported here are for stationary and 
low Reynolds number flow. It thus differs from \cite{Ch04},
which contains a significant time dependence.
The interface shape 
(cf. Fig.~\ref{fig:image}) is recorded using a microscope with a 
working distance of $20$ mm and a a high resolution CCD camera. 
The overall resolution of the optics is 1 $\rm{\mu m}$.
The benefit of using gravity to drive the flow is that no 
unsteadiness could be detected down to a micron scale, but
we are limited in the achievable flow rates. Our attempts of using
a pump for driving failed owing to unsteadiness of the interface caused 
by mechanical vibrations.

\begin{figure}
\includegraphics[width=8cm]{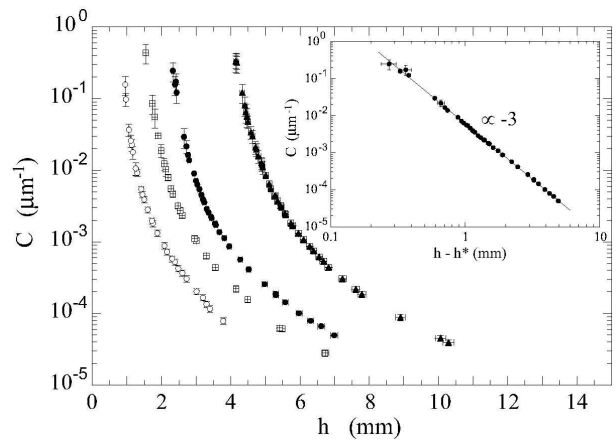}
\caption{Tip curvature $C$ as a function of $h$ for silicone oil
($\eta$ = 30 Pa s, $\gamma\rm{= 2.13\cdot 1.0\times 10^{-2}N m^{-1}}$), 
$a$ = 1 mm, and $Q$=1.7 $10^{-3}$ ml s$^{-1}$($\circ$), $5.1\cdot 10^{-3} 
\rm{ml s^{-1}}$ ($\boxplus$), $10^{-2} \rm{ml s^{-1}}$
($\bullet$) and for $a$ = 3 mm and $Q$ = $3.9\cdot10^{-2}
\rm{ml s^{-1}}$ ($\blacktriangle$). Errors in the curvature 
represent typical deviations between Gaussian and parabolic fits,
whereas errors of 1.5 $\%$ in $h$ come from the finite range of focus.
Inset: log-log plot of $C$ versus $h-h^{\star}$ for 
$Q =10^{-2}\rm{ml s^{-1}}$.\label{fig:Ch}}
\end{figure}

Back-lighting ensures a sharp gray-level gradient at the interface, 
and profiles $R(z)$ were extracted in numerical form using image
processing software. Axisymmetry is preserved to within 
our experimental resolution.
First we focus on the development of a sharp
tip, shown in Fig.~\ref{fig:image} a),b). To that end we determine 
the curvature $C$ at the tip, by fitting a 
Gaussian and a parabolic profile (cf. Fig.~\ref{fig:image} d)) to it. 
Fig.~\ref{fig:Ch} shows the curvature as a function of the distance
$h$ from the orifice to the tip for various flow rates $Q$, using the same
silicone oil of viscosity $\eta=30$ Pa s, surface tension 
$\gamma\rm{= 2.13\cdot 10^{-2} N m^{-1}}$, and density 
$\rho=976 kg/m^3$. As $h$ reaches 
a {\it critical} value $h^{\star}$, the curvature diverges like
$1/(h-h^{\star})^3$, as shown in the inset for one data set. 
The same exponent $-3\pm0.2$ is observed for all our data sets.
Albeit for a limited range, the data of \cite{Ch04} for $H$ is 
also consistent with ours. 

\begin{figure}
\includegraphics[width=8cm]{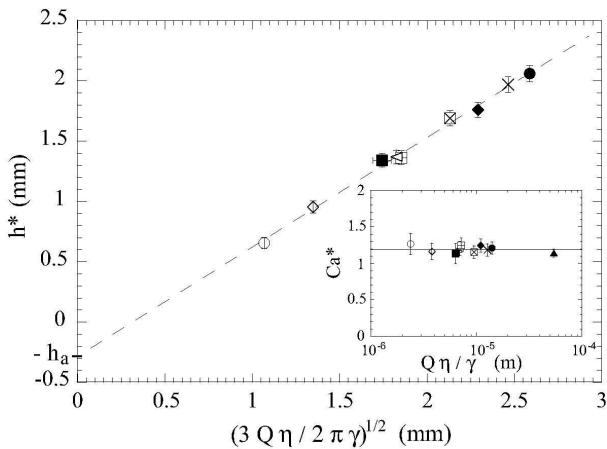}
\caption{Critical $h=h^{\star}$ with $a$ = 1 mm for silicone oil
($\eta$=30 Pa s) and $Q/\rm{ml s^{-1}}$=1.7 $\cdot 10^{-3}$($\circ$), 
$5.1\cdot 10^{-3}$($\boxplus$), $6.8\cdot 10^{-3}$ ($\boxtimes$), 
$9\cdot 10^{-3}$ ($\times$), $10^{-2}$ ($\bullet$), 
silicone oil ($\eta$=60 Pa s) and $Q/\rm{ml s^{-1}}$=
1.4 $\cdot 10^{-3}$($\diamond$), 2.5$\cdot 10^{-3}$($\triangleleft$),
3.9 $\cdot 10^{-3}$($\blacklozenge$), castor oil and 
$Q/\rm{ml s^{-1}}$= 0.23 ($\blacksquare$). Errors in
$h^{\star}$ come from the absolute measurement of the orifice position,
parameters for castor oil: $\gamma=3.53\pm0.1 \cdot 10^{-2} \rm{N m^{-1}}, 
\eta=0.93\pm0.01 \rm{Pa s}$. Inset: capillary number $Ca^{\star}$
for all experiments (same symbols, $\blacktriangle$: $a$ = 3 mm 
with silicone oil ($\eta$= 30 Pa s) and 
$Q/\rm{ml s^{-1}}$ = $3.9\cdot 10^{-2}$.) 
\label{fig:hStar}}
\end{figure}

We now show that the singularity occurs when the tip experiences a critical 
viscous forcing, relative to the smoothing effects of surface tension.
Namely, in the limit of slow flow a point sink in a solid plate 
produces a velocity  $3Q /(2 \pi z^2)$ along the line of symmetry
\cite{HB65}, p.140. Thus equating viscous forcing by the 
{\it unperturbed} flow field with surface tension leads to 
$3\eta Q/(2 \pi {h^{\star}}^{2}) \sim \gamma$ for the distance
$h^{\star}$ where the singularity is expected to occur. This is consistent
with the data presented in Fig.~\ref{fig:hStar}, if we assume that the 
point sink is placed a small distance $h_a=0.29a$ {\it inside} the 
hole of diameter $a=1$mm. Thus the singularity is characterized 
by the {\it local} capillary number 
\begin{equation}
Ca(h) = \frac{3Q\eta}{2\pi(h+h_a)^{2}\gamma}
\label{Ca}
\end{equation}
reaching a critical value
\begin{equation}
Ca(h^{\star}) \equiv Ca^{\star} = 1.2\pm 0.06.
\label{Cacr}
\end{equation}
This is confirmed in the inset of Fig.~\ref{fig:hStar} using all our data sets,
obtained with two different tube openings $a=1$mm and $a=3$mm.
For the shift we used $h_a=0.29a$, a form consistent with a transition
from a sink to a Poiseuille flow \cite{H04}.

\begin{figure}
\includegraphics[width=8cm]{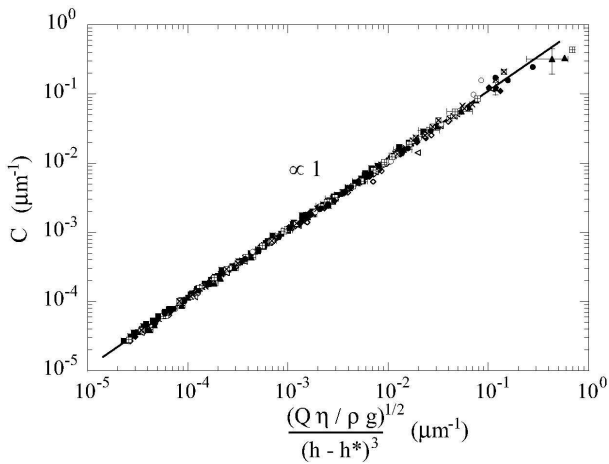}
\caption{Master curve, collapsing all data according to (\ref{curvature}).
Symbols are the same than in Fig.~\ref{fig:hStar}.
Castor oil data with $C^{-1} \le 25\mu m$ are not shown, owing to
dust perturbing the tip.\label{fig:Master}}.
\end{figure}

Utilizing the experimental observation that the {\it prefactor} of 
the curvature scales like $\sqrt{Q}$ for a given liquid, we can 
collapse {\it all} the curvature data into a single scaling law:
\begin{equation}
C=(1.08\pm0.06)\frac{[Q\eta/(\rho g)]^{1/2}}{(h-h^{\star})^3}=
1.88\frac{\ell_c(h^{\star}+h_a)}{(h-h^{\star})^3}, 
\label{curvature}
\end{equation}
where $\ell_c=\sqrt{\gamma/(\rho g)}$ is the capillary length. 
Figure~\ref{fig:Master} shows the master curve corresponding 
to (\ref{curvature}) for all our experiments, covering four
orders of magnitude in the curvature. As expected, (\ref{curvature})
says that the curvature increases with flow rate, while gravity tends
to flatten the interface. Surface tension is contained implicitly 
through the dependence on $h^{\star}$, given by (\ref{Cacr}).
It is instructive to also split the r.h.s. of (\ref{curvature})
into dependencies on the ``internal'' variable $h$ and on the ``external''
capillary length $\ell_c$, which is required on dimensional grounds. 
Silicone oil and and castor oil have capillary lengths $\ell_c$ 
of 1.49 mm and 1.94 mm respectively, so we have good experimental 
evidence that it is indeed $\ell_c$ that sets the external length 
scale in (\ref{curvature}), rather than the size of the cell. 

We now turn to the interface shape for $h < h^{\star}$, for which 
the tip is singular, and the profile meets the tip with a finite 
slope $R'(h)$. As the hole is approached, the slope becomes small, 
suggesting the use of slender-body theory \cite{T66,B72,AL78}. 
In this limit, the perturbation of the axial flow by the interface 
is small, so it can be represented as a line distribution of 2D 
sources \cite{AL78}. For the slope of the interface at the tip we thus find 
\begin{equation}
R'(h) = 1/[2Ca(h)],
\label{slope}
\end{equation}
where $Ca$ is given by (\ref{Ca}). Qualitatively, (\ref{slope})
results from the ratio of the unperturbed flow speed in the 
$z$-direction and the capillary speed $\gamma/\eta$ contracting
the profile in the $r$-direction, while gravity is irrelevant near the tip.
\begin{figure}
\includegraphics[width=8cm]{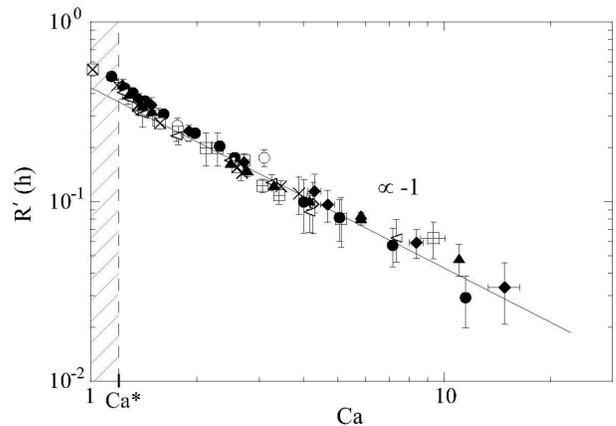}
\caption{Slope $R'(h)$ versus $Ca$ for all experiments but the 
one with castor oil. Same symbols as in Fig.~\ref{fig:hStar}. 
Line: best fit with $R'(h)=(0.43\pm0.1)/Ca$. In the shaded region
the slope is defined as the minimum value of $R'(z)$.
\label{fig:Pente}}
\end{figure}

The slope $R'$ was determined experimentally
by fitting $R(z)$ with a second order polynomial in $|z-h|$
and extrapolating to the tip. The result, shown in Fig.~\ref{fig:Pente},
is in excellent agreement with the predicted slope, in particular
for small opening angles, as expected. The prefactor is
$0.43\pm0.1$, only slightly smaller than the predicted 
value of $0.5$. To estimate the local capillary number, we included
the correction $h_a$ deduced from $h^{\star}$. 
This we believe to be the first quantitative test 
of this classical slender body theory, previous comparisons 
being only qualitative \cite{BL86}. The shaded region of Fig.~\ref{fig:Pente}
corresponds to {\it smooth} profiles below the critical capillary
number, where we took the minimum value of $R'(z)$ as the slope. 
Its value merges smoothly into the singular region. At the transition 
$Ca=Ca^{\star}$, the interface slope assumed a universal value of 
$R'(h^{\star})=0.47\pm0.06$ for all our experiments.

{\it Discussion:} Our study was motivated by the aforementioned
viscous withdrawal experiment \cite{CN02,C04}, which found the 
tip to disappear at a typical size of 200 $\mu$m, independent 
of the viscosity ratio $\lambda$ between the two liquids. 
This differs markedly from both theory \cite{AL78} and 
experiment \cite{BL86} with drops of small viscosity in strong flows, 
which develop tips which are locally similar, but which remain 
stable well into the singular regime. We thus expected the 
disappearance of the tip \cite{CN02} to be related to the far-field 
boundary condition, which is a flat interface for viscous withdrawal,
but a curved interface for drops. 

However, at the resolution of the present 
experiment ($\approx 1 \rm{\mu m}$), no air entrainment was ever observed.
In fact, air bubbles suspended above the sink hole \cite{CE04}
give results very similar to the ones described here, except that 
the curvature scaling does not extend to as small values, owing to
the bubble curvature. We have considered a number of possible causes
for the liquid-liquid and the liquid-air system to behave differently,
yet were not able to single out a likely mechanism:

a) In \cite{C04} it was proposed that the minimum tip size behaves 
approximately as $C_{min}^{-1}\sim0.1\sqrt{\gamma/(\Delta\rho g)}$, owing to
the stabilizing effect of the density difference $\Delta\rho$ 
between the two liquids. However, this gives a minimum tip size of 
150$\mu$m for a liquid-air experiment, in disagreement with our observations.
b) Our viscosity ratios (between 
$\lambda\equiv\eta_{air}/\eta_{liquid}=3\cdot 10^{-7}$ 
and $2\cdot 10^{-5}$) are smaller than those of \cite{C04}. 
However, in neither experiment could a significant dependence 
on $\lambda$ be detected. 
c) Surfactants are known to lead to the ejection of a 
jet out of the tip of bubbles \cite{B93} (``tip streaming''). 
However, care was taken in \cite{C04} to differentiate this
transient phenomenon from the reported instability.
d) Finally, it remains to mention the possible effects of 
compressibility, solubility, and of evaporation into the gas,
which distinguish a gas most markedly from a liquid.
In fact solubility has, in the two-dimensional analogue of the 
present problem, been suggested \cite{J02} as a mechanism to relieve 
the pressure that builds up inside a cusp, and which leads to its
eventual disappearance \cite{E01}. However, one will require
a better knowledge of the flow near the tip to estimate the correct 
pressure balance in the present axisymmetric case. It should also 
be noted that once a tip size of 1 micron is reached, the mean
free path of the air will come into play.

As long as the tip is above the orifice, no air entrainment occurs. 
However the indirect evidence of air bubbles appearing at the other
end of the tube shows that a spout is formed once the tip enters 
the orifice. In fact, this setup has been used to 
create monodisperse streams of bubbles \cite{GG01}. 
When subsequently the filling height $H$ is increased
again to raise the tip above the orifice, no spout is observed.
Rather, a stable tip reappears at $h=0$,
giving no evidence for hysteresis. Thus the experimental 
protocol suggested in \cite{Z04} fails to produce a thin spout, 
a fact that could be related to our far-field conditions being
different from those of \cite{Z04}.

A theoretical understanding of the scaling law (\ref{curvature})
remains a challenge. Likely, such an understanding will require the 
description of the entire interface profile. Rescaling the profiles
using $C^{-1}$ as a characteristic scale has lead to data collapse 
only in a very limited region around the tip. Evidently
the profiles contain several length scales that need to be 
included into a proper description. 

In conclusion, we have developed a scaling law for 
the formation of an axisymmetric tip singularity on a 
liquid-air interface. Singular tips are well described by
slender-body theory. Our analysis raises the possibility of
using hydrodynamics to produce structures of virtually 
unlimited smallness. 

\acknowledgments
We are grateful to R. Deegan for his help in designing
the experiment and his continued support, and to I. Cohen,
R. Kerswell, and \'E. Lorenceau for very helpful discussions.

\bibliography{revise}

\begin{thebibliography}{27}
\expandafter\ifx\csname natexlab\endcsname\relax\def\natexlab#1{#1}\fi
\expandafter\ifx\csname bibnamefont\endcsname\relax
  \def\bibnamefont#1{#1}\fi
\expandafter\ifx\csname bibfnamefont\endcsname\relax
  \def\bibfnamefont#1{#1}\fi
\expandafter\ifx\csname citenamefont\endcsname\relax
  \def\citenamefont#1{#1}\fi
\expandafter\ifx\csname url\endcsname\relax
  \def\url#1{\texttt{#1}}\fi
\expandafter\ifx\csname urlprefix\endcsname\relax\def\urlprefix{URL }\fi
\providecommand{\bibinfo}[2]{#2}
\providecommand{\eprint}[2][]{\url{#2}}

\bibitem[{\citenamefont{Eggers}(1993)}]{E93}
\bibinfo{author}{\bibfnamefont{J.}~\bibnamefont{Eggers}},
  \bibinfo{journal}{Phys. Rev. Lett.} \textbf{\bibinfo{volume}{71}},
  \bibinfo{pages}{3458} (\bibinfo{year}{1993}).

\bibitem[{\citenamefont{Rothert et~al.}(2001)\citenamefont{Rothert, Richter,
  and Rehberg}}]{RRR01}
\bibinfo{author}{\bibfnamefont{A.}~\bibnamefont{Rothert}},
  \bibinfo{author}{\bibfnamefont{R.}~\bibnamefont{Richter}}, \bibnamefont{and}
  \bibinfo{author}{\bibfnamefont{I.}~\bibnamefont{Rehberg}},
  \bibinfo{journal}{Phys. Rev. Lett.} \textbf{\bibinfo{volume}{87}},
  \bibinfo{pages}{084501} (\bibinfo{year}{2001}).

\bibitem[{\citenamefont{Eggers et~al.}(1999)\citenamefont{Eggers, Lister, and
  Stone}}]{ELS99}
\bibinfo{author}{\bibfnamefont{J.}~\bibnamefont{Eggers}},
  \bibinfo{author}{\bibfnamefont{J.}~\bibnamefont{Lister}}, \bibnamefont{and}
  \bibinfo{author}{\bibfnamefont{H.}~\bibnamefont{Stone}}, \bibinfo{journal}{J.
  Fluid Mech.} \textbf{\bibinfo{volume}{401}}, \bibinfo{pages}{293}
  (\bibinfo{year}{1999}).

\bibitem[{\citenamefont{Landau and Lifshitz}(1984)}]{LL84}
\bibinfo{author}{\bibfnamefont{L.}~\bibnamefont{Landau}} \bibnamefont{and}
  \bibinfo{author}{\bibfnamefont{E.}~\bibnamefont{Lifshitz}},
  \emph{\bibinfo{title}{Fluid Mechanics}} (\bibinfo{publisher}{Pergamon,
  Oxford}, \bibinfo{year}{1984}).

\bibitem[{\citenamefont{Eggers}(2001)}]{E01}
\bibinfo{author}{\bibfnamefont{J.}~\bibnamefont{Eggers}},
  \bibinfo{journal}{Phys. Rev. Lett.} \textbf{\bibinfo{volume}{86}},
  \bibinfo{pages}{4290} (\bibinfo{year}{2001}).

\bibitem[{\citenamefont{Lorenceau et~al.}(2003)\citenamefont{Lorenceau,
  Restagno, and Qu\'{e}r\'{e}}}]{LRQ03}
\bibinfo{author}{\bibfnamefont{E.}~\bibnamefont{Lorenceau}},
  \bibinfo{author}{\bibfnamefont{F.}~\bibnamefont{Restagno}}, \bibnamefont{and}
  \bibinfo{author}{\bibfnamefont{D.}~\bibnamefont{Qu\'{e}r\'{e}}},
  \bibinfo{journal}{Phys.\ Rev. \ Lett.} \textbf{\bibinfo{volume}{90}},
  \bibinfo{pages}{184501} (\bibinfo{year}{2003}).

\bibitem[{\citenamefont{Taylor}(1934)}]{T34}
\bibinfo{author}{\bibfnamefont{G.}~\bibnamefont{Taylor}},
  \bibinfo{journal}{Proc. R. Soc. London A} \textbf{\bibinfo{volume}{146}},
  \bibinfo{pages}{501} (\bibinfo{year}{1934}).

\bibitem[{\citenamefont{Cohen and Nagel}(2002)}]{CN02}
\bibinfo{author}{\bibfnamefont{I.}~\bibnamefont{Cohen}} \bibnamefont{and}
  \bibinfo{author}{\bibfnamefont{S.~R.} \bibnamefont{Nagel}},
  \bibinfo{journal}{Phys.\ Rev. \ Lett.} \textbf{\bibinfo{volume}{88}},
  \bibinfo{pages}{074501} (\bibinfo{year}{2002}).

\bibitem[{\citenamefont{Chaieb}(2004)}]{Ch04}
\bibinfo{author}{\bibfnamefont{S.}~\bibnamefont{Chaieb}},
  \bibinfo{journal}{arXiv:physics/0404088}  (\bibinfo{year}{2004}).

\bibitem[{\citenamefont{Eggers}(1998)}]{E98}
\bibinfo{author}{\bibfnamefont{J.}~\bibnamefont{Eggers}}, in
  \emph{\bibinfo{booktitle}{A Perspective Look at Nonlinear Media in Physics,
  Chemistry, and Biology}}, edited by
  \bibinfo{editor}{\bibfnamefont{J.}~\bibnamefont{Parisi}},
  \bibinfo{editor}{\bibfnamefont{S.~C.} \bibnamefont{Mueller}},
  \bibnamefont{and}
  \bibinfo{editor}{\bibfnamefont{W.}~\bibnamefont{Zimmermann}}
  (\bibinfo{publisher}{Springer, Berlin}, \bibinfo{year}{1998}), pp.
  \bibinfo{pages}{305--312}.

\bibitem[{\citenamefont{Kadanoff}(1997)}]{K97}
\bibinfo{author}{\bibfnamefont{L.}~\bibnamefont{Kadanoff}},
  \bibinfo{journal}{Phys. Today} \textbf{\bibinfo{volume}{50(9)}},
  \bibinfo{pages}{11} (\bibinfo{year}{1997}).

\bibitem[{\citenamefont{Jeong and Moffatt}(1992)}]{JM92}
\bibinfo{author}{\bibfnamefont{J.-T.} \bibnamefont{Jeong}} \bibnamefont{and}
  \bibinfo{author}{\bibfnamefont{H.}~\bibnamefont{Moffatt}},
  \bibinfo{journal}{J. Fluid Mech.} \textbf{\bibinfo{volume}{241}},
  \bibinfo{pages}{1} (\bibinfo{year}{1992}).

\bibitem[{\citenamefont{Stone}(1994)}]{S94}
\bibinfo{author}{\bibfnamefont{H.}~\bibnamefont{Stone}},
  \bibinfo{journal}{Annu. Rev. Fluid Mech.} \textbf{\bibinfo{volume}{26}},
  \bibinfo{pages}{65} (\bibinfo{year}{1994}).

\bibitem[{\citenamefont{Ga{\~n}{\'a}n-Calvo}(1998)}]{G98}
\bibinfo{author}{\bibfnamefont{A.}~\bibnamefont{Ga{\~n}{\'a}n-Calvo}},
  \bibinfo{journal}{Phys. Rev. Lett.} \textbf{\bibinfo{volume}{80}},
  \bibinfo{pages}{285} (\bibinfo{year}{1998}).

\bibitem[{\citenamefont{Cohen}(2004)}]{C04}
\bibinfo{author}{\bibfnamefont{I.}~\bibnamefont{Cohen}},
  \bibinfo{journal}{Phys.\ Rev. E} \textbf{\bibinfo{volume}{70}},
  \bibinfo{pages}{026302} (\bibinfo{year}{2004}).

\bibitem[{\citenamefont{Ga{\~n}{\'a}n-Calvo and Gordillo}(2001)}]{GG01}
\bibinfo{author}{\bibfnamefont{A.}~\bibnamefont{Ga{\~n}{\'a}n-Calvo}}
  \bibnamefont{and} \bibinfo{author}{\bibfnamefont{J.}~\bibnamefont{Gordillo}},
  \bibinfo{journal}{Phys. Rev. Lett.} \textbf{\bibinfo{volume}{87}},
  \bibinfo{pages}{274501} (\bibinfo{year}{2001}).

\bibitem[{\citenamefont{Stone et~al.}(2004)\citenamefont{Stone, Stroock, and
  Ajdari}}]{SSA04}
\bibinfo{author}{\bibfnamefont{H.}~\bibnamefont{Stone}},
  \bibinfo{author}{\bibfnamefont{A.}~\bibnamefont{Stroock}}, \bibnamefont{and}
  \bibinfo{author}{\bibfnamefont{A.}~\bibnamefont{Ajdari}},
  \bibinfo{journal}{Annu. Rev. Fluid Mech.} \textbf{\bibinfo{volume}{36}},
  \bibinfo{pages}{381} (\bibinfo{year}{2004}).

\bibitem[{\citenamefont{Zhang}(2004)}]{Z04}
\bibinfo{author}{\bibfnamefont{W.}~\bibnamefont{Zhang}},
  \bibinfo{journal}{Phys. Rev. Lett.} \textbf{\bibinfo{volume}{93}},
  \bibinfo{pages}{184502} (\bibinfo{year}{2004}).

\bibitem[{\citenamefont{Happel and Brenner}(1965)}]{HB65}
\bibinfo{author}{\bibfnamefont{J.}~\bibnamefont{Happel}} \bibnamefont{and}
  \bibinfo{author}{\bibfnamefont{H.}~\bibnamefont{Brenner}},
  \emph{\bibinfo{title}{Low Reynolds Number Hydrodynamics}}
  (\bibinfo{publisher}{Prentice-Hall, Englewood Cliffs, NJ},
  \bibinfo{year}{1965}).

\bibitem[{\citenamefont{Hinch}(2004)}]{H04}
\bibinfo{author}{\bibfnamefont{E.}~\bibnamefont{Hinch}},
  \bibinfo{journal}{private communication}  (\bibinfo{year}{2004}).

\bibitem[{\citenamefont{Taylor}(1966)}]{T66}
\bibinfo{author}{\bibfnamefont{G.}~\bibnamefont{Taylor}}, in
  \emph{\bibinfo{booktitle}{Proceedings of the 11th International Congress of
  Applied Mathematics}}, edited by
  \bibinfo{editor}{\bibfnamefont{G.}~\bibnamefont{Batchelor}}
  (\bibinfo{publisher}{Springer, Heidelberg}, \bibinfo{year}{1966}).

\bibitem[{\citenamefont{Buckmaster}(1972)}]{B72}
\bibinfo{author}{\bibfnamefont{J.}~\bibnamefont{Buckmaster}},
  \bibinfo{journal}{J. Fluid Mech.} \textbf{\bibinfo{volume}{55}},
  \bibinfo{pages}{385} (\bibinfo{year}{1972}).

\bibitem[{\citenamefont{Acrivos and Lo}(1978)}]{AL78}
\bibinfo{author}{\bibfnamefont{A.}~\bibnamefont{Acrivos}} \bibnamefont{and}
  \bibinfo{author}{\bibfnamefont{T.}~\bibnamefont{Lo}}, \bibinfo{journal}{J.
  Fluid Mech.} \textbf{\bibinfo{volume}{86}}, \bibinfo{pages}{641}
  (\bibinfo{year}{1978}).

\bibitem[{\citenamefont{Bentley and Leal}(1986)}]{BL86}
\bibinfo{author}{\bibfnamefont{B.}~\bibnamefont{Bentley}} \bibnamefont{and}
  \bibinfo{author}{\bibfnamefont{L.}~\bibnamefont{Leal}}, \bibinfo{journal}{J.
  Fluid Mech.} \textbf{\bibinfo{volume}{167}}, \bibinfo{pages}{219}
  (\bibinfo{year}{1986}).

\bibitem[{\citenamefont{{Courrech du Pont} and Eggers}(2004)}]{CE04}
\bibinfo{author}{\bibfnamefont{S.}~\bibnamefont{{Courrech du Pont}}}
  \bibnamefont{and} \bibinfo{author}{\bibfnamefont{J.}~\bibnamefont{Eggers}},
  \bibinfo{journal}{unpublished}  (\bibinfo{year}{2004}).

\bibitem[{\citenamefont{Bruijn}(1993)}]{B93}
\bibinfo{author}{\bibfnamefont{R.~D.} \bibnamefont{Bruijn}},
  \bibinfo{journal}{Chem. Eng. Sci.} \textbf{\bibinfo{volume}{48}},
  \bibinfo{pages}{277} (\bibinfo{year}{1993}).

\bibitem[{\citenamefont{Jacqmin}(2002)}]{J02}
\bibinfo{author}{\bibfnamefont{D.}~\bibnamefont{Jacqmin}}, \bibinfo{journal}{J.
  Fluid Mech.} \textbf{\bibinfo{volume}{455}}, \bibinfo{pages}{347}
  (\bibinfo{year}{2002}).

\end{thebibliography}

\end{document}